\documentclass[pdftex,aipproc]{revtex4-1}
\usepackage{graphicx}
\usepackage{amsmath}
\usepackage{graphicx}

\begin{document}
\title{What does it mean to have `seen' the quark-gluon plasma?}
\author{Scott Pratt}
\affiliation{Department of Physics and Astronomy and National Superconducting Cyclotron Laboratory,
Michigan State University\\
East Lansing, Michigan 48824}
\date{\today}
\begin{abstract}
Identifying the quark-gluon plasma requires convincing experimental evidence that partons move independently throughout the environment created in a heavy ion collision and with densities expected from equilibrium considerations. In lattice calculations, charge correlations suggest that quarks exist independently, and are not merely exchanged from hadronic object to another. Many experimental signatures (J/Psi suppression, quark number scaling, etc.) suggest that quarks are not confined to their original singlets, but these signatures do not make a clear case that quarks move independently or that they have the expected densities. I discuss a class of measurements that parallel lattice observables and has the prospect of investigating whether partonic charges move independently.
\end{abstract}

\maketitle

Lattice calculations show that the microscopic degrees of freedom undergo a dramatic change from hadrons to partons in a temperature window around 175 MeV. In these calculations charge fluctuations have been extracted as a three-by-three matrix, and expressed as a function of temperature \cite{Borsanyi:2010cj,Bazavov:2012jq}
\begin{equation}
\chi_{ab}=\langle Q_aQ_b\rangle/V,
\end{equation}
where $V$ is the volume and the indices $a$ and $b$ refer to the net amount of up, down and strange charge. For the purposes of this talk, I consider only neutral systems, and ignore the subtraction of the terms $\langle Q_a\rangle\langle Q_b\rangle$. Charge correlations provide insight into the microscopic degrees of freedom. For a gas of uncorrelated particles, the only correlations are between the charges in the same particle. For a partonic gas the quasi-particles each have a single charge and,
\begin{equation}
\label{eq:chihad}
\chi_{ab}^{\rm(QGP)}=2\delta_{ab}n_{a},
\end{equation}
where $n_a$ is the density of up, down or strange quarks. In contrast, hadrons have multiple charges and for a hadron gas,
\begin{equation}
\chi_{ab}^{\rm(HAD)}=\sum_\alpha n_\alpha Q_{\alpha,a}Q_{\alpha,b},
\end{equation}
where $Q_{\alpha,a}$ is the net number of $a$ quarks in a hadron species $\alpha$. Unlike the parton gas, the charge fluctuations for a hadron gas have off-diagonal components \cite{Koch:2005vg}.

\begin{figure}[b!]
\centerline{\includegraphics[width=0.4\columnwidth]{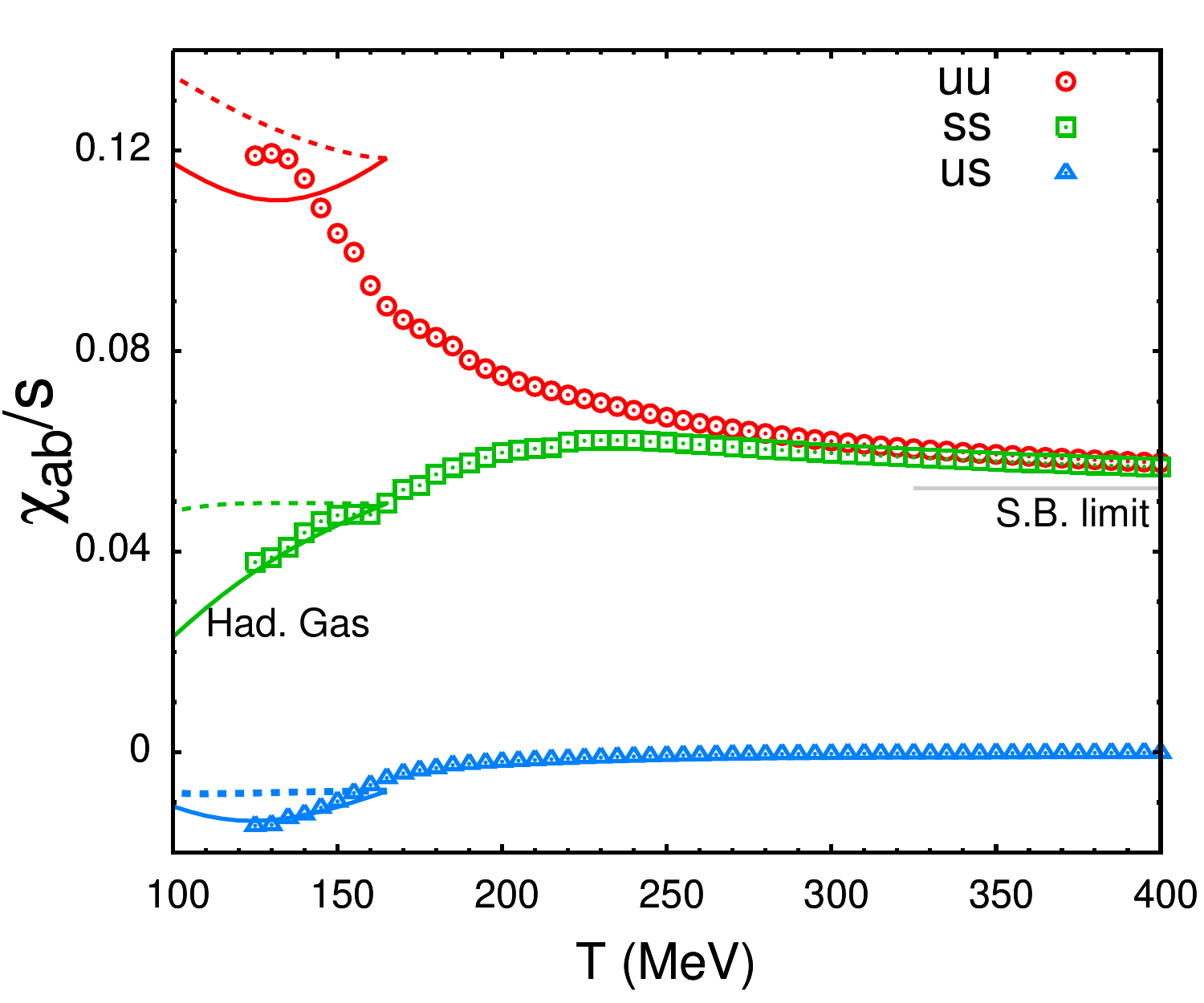}}
\caption{\label{fig:claudia}
Charge correlations from lattice calculations are shown as a function of the temperature. At high temperature, the correlations approach the Stefan-Boltzmann limit of a parton gas, while at low temperature the result is consistent with a hadron gas (solid line). The dashed line shows the correlations expected if a chemically equilibrated gas at $T=165$ is allowed to lose chemical equilibrium according to chemical evolution equations. Whereas $\chi_{uu}$ roughly doubles during the hadronization region, $\chi_{ss}$ is modestly reduced.}
\end{figure}
Figure $\ref{fig:claudia}$ shows the ratio $\chi/s$ as a function of temperature as calculated by the Wuppertal-Budapest lattice group \cite{Borsanyi:2010cj}. At high temperature the ratio approaches expectations for the Stefan-Boltzmann limit for a parton gas, while at low temperatures it approaches that of a hadron gas. As expected, the off-diagonal elements disappear above $T_c$. A second feature of the behavior is that $\chi_{uu}/s$ and $\chi_{dd}/s$ rise sharply as the temperature enters the critical region, while $\chi_{ss}/s$ stays roughly constant. To understand this, one can consider the isentropic expansion of a volume which initially has 1000 partons. Roughly two thirds of those hadrons might be quarks, while the others are gluons. In order to conserve entropy, the 1000 partons must hadronize into roughly 1000 hadrons. But since each hadron is composed of at least two quarks, the number of quarks should more than double at hadronization. Additionally, since strange quarks are suppressed in a hadron gas due to the relatively large mass of kaons, the net number of strange quarks should stay roughly the same.

Unfortunately, the charge correlations in Fig. \ref{fig:claudia} cannot be directly compared to experimental data.  The main difficulty derives from the fact that charge is locally conserved in a collision, and the net charge does not fluctuate. However, the charge does diffuse a finite distance, and the charge-charge correlation matrix is driven by the quantities $\chi_{ab}$. This paper employs the methods described in \cite{Pratt:2012dz} which gives the means to overcome these challenges. We consider the charge-charge correlation matrix,
\begin{equation}
g_{ab}(\Delta\eta)\equiv
\langle\rho_a(0)\rho_b(\Delta\eta)\rangle,
\end{equation}
where $\rho_a(\eta)$ is the net density per unit spatial rapidity of charges of type $a$. Assuming boost invariance, $g_{ab}$ depends only on the rapidity difference $\Delta\eta$. If the correlation is measured at breakup density, and if the correlation includes only that between different particles, charge conservation demands 
\begin{equation}
\int d\Delta\eta~g_{ab}(\Delta\eta)=-\chi_{ab}^{\rm(HAD)}.
\end{equation}
where $\chi^{\rm(HAD)}$ is defined as in Eq. (\ref{eq:chihad}) except with $n_\alpha$ referring to densities per unit spatial rapidity, a quantity which can be taken from final-state yields. A second constraint on $\chi$ is that the long-range part should be dominated by the its structure in the QGP. This comes from the fact that after an initial burst of charge creation during the thermalization of the QGP, the quark number is roughly fixed. The balancing charges should then diffuse away from one another and produce a gaussian-like structure for $g_{ab}(\Delta\eta)$, that integrates to $-\chi_{ab}^{\rm(QGP)}$. Due to local charge conservation, the second wave of creation and annihilation processes near $T_c$ should only alter the short-scale behavior of $g_{ab}(\Delta\eta)$, while the long scale behavior should simply continue to diffuse. By making an oversimplification that charge creation has two scales (not true since both waves of charge production are not instantaneous), one can express $g_{ab}(\Delta\eta)$ as
\begin{equation}
\label{eq:gab2wave}
g_{ab}(\Delta\eta)=-\left(\chi_{ab}^{\rm(HAD)}-\chi_{ab}^{\rm(QGP)}\right)\frac{e^{-(\Delta\eta)^2/2\sigma_{\rm(HAD)}^2}}{(2\pi\sigma_{\rm(HAD)})^{1/2}}
-\chi_{ab}^{\rm(QGP)}\frac{e^{-(\Delta\eta)^2/2\sigma_{\rm(QGP)}^2}}{(2\pi\sigma_{\rm(QGP)})^{1/2}}.
\end{equation}
From rough estimates, one might expect the diffusive widths to be $\approx 1$ unit for $\sigma_{\rm(QGP)}$ and only a few tenths of a unit of rapidity for $\sigma_{\rm(HAD)}$. 

More generally, one could consider the case for continuous creation or annihilation of charge. In that case,
\begin{equation}
\label{eq:realthing}
g_{ab}(\Delta\eta,\tau)=-\int_0^\tau d\tau'~
\frac{d\chi_{ab}(\tau')}{d\tau'}D(\Delta\eta,\tau',\tau),
\end{equation}
where $D(\Delta\eta,\tau,\tau')$ describes the normalized probability of charges diffusing away from one another a distance $\Delta\eta$ at time $\tau$ given they were created at $\tau'$. This more general form could truthfully incorporate the expected rate of charge creation shown in the lattice calculations of Fig. \ref{fig:claudia}. For this study, we consider the simple two-wave picture. 

The next roadblock to overcome in comparing to experiment is that hadrons are measured, not charge. It was shown in \cite{Pratt:2012dz} that the hadronic correlator,
\begin{equation}
g_{\alpha\beta}(\Delta\eta)=\langle(n_{\alpha}(0)-n_{\bar{\alpha}})(n_{\beta}(\Delta\eta)-n_{\bar{\beta}}(\Delta\eta)\rangle.
\end{equation}
could be generated from $g_{ab}$ by assuming balancing charges were spread amongst the species thermally. To apply the thermal weights, one associates a Lagrange multiplier, $\mu_{ab}(\Delta\eta)$ to the two-particle weights, i.e.,
\begin{equation}
\langle n_\alpha(0)n_\beta(\Delta\eta)\rangle=
\langle n_\alpha\rangle\langle n_\beta\rangle e^{\mu_{ab}(\Delta\eta)Q_{\alpha,a}Q_{\beta,b}}.
\end{equation}
Since the correlations are small, $\mu$ is small and one can solve for the final hadronic correlators \cite{Pratt:2012dz},
\begin{equation}
\label{eq:galphabeta}
g_{\alpha\beta}(\Delta\eta)=-2\sum_{abcd}Q_{\alpha,a}\chi_{ab}^{-1\rm(HAD)}g_{bc}(\Delta\eta)
\chi_{cd}^{-1\rm(HAD)}Q_{\beta,d}.
\end{equation}

\begin{figure}
\centerline{\includegraphics[width=0.4\columnwidth]{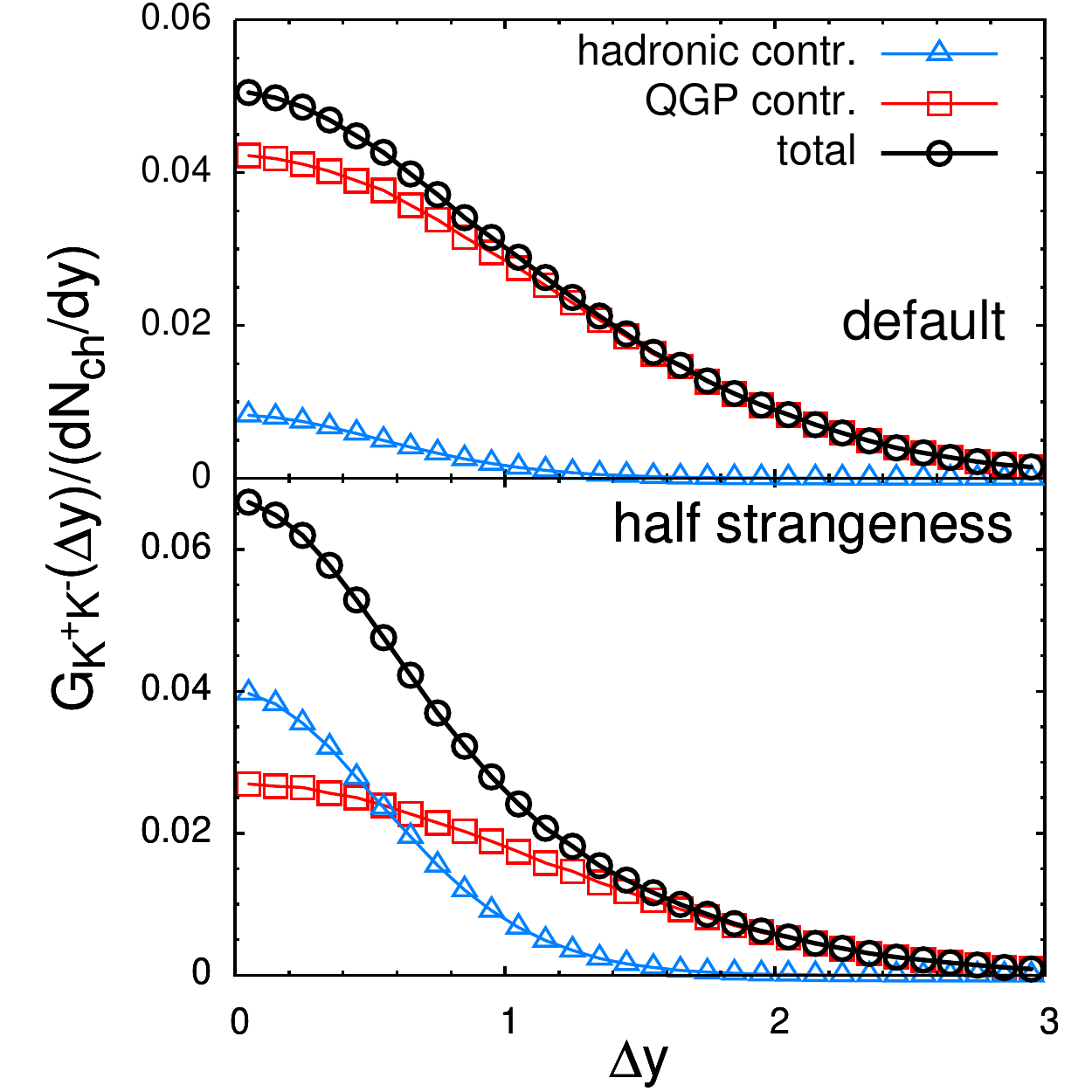}}
\caption{\label{fig:GKK}
The $K^+K^-$ correlator is calculated assuming that the strangeness is fully equilibrated in the QGP (upper panel) and assuming half the strangeness content in the lower panel. In the fully equilibrated calculation, the correlator is dominated by the long-range contribution as described in Eq. (\ref{eq:gab2wave}). For a strangeness depleted plasma, one needs more of the strange quarks to be produced at hadronization, which strengthens the narrow contribution and leads to a noticeably narrow correlation. Details of the calculation can be found in \cite{Pratt:2012dz}.}
\end{figure}

Combining Eq.s (\ref{eq:gab2wave}) and (\ref{eq:galphabeta}) gives the correlation in coordinate space between any two hadronic species. Convoluting the correlation in relative spatial rapidity with a thermal blast wave model \cite{Retiere:2003kf} and a Monte Carlo of the decays gives the final correlation in coordinate space \cite{Pratt:2012dz}. Figure \ref{fig:GKK} shows $G_{K^+K^-}$ for two cases. In each case, the diffusive widths are $\sigma_{\rm(QGP)}=1, \sigma_{\rm(HAD)}=0.2$. In the upper panel, the density of strange quarks in the QGP is assumed to be the equilibrium value, while in the lower panel the strange quark density is assumed to be half the equilibrium value. By halving the density in the QGP, the strength of the narrow component rises to produce sufficient strangeness for the final state. By strengthening the narrow component and weakening the broader component, the resulting balance function would be much narrower. Thus, measuring the $K^+K^-$ correlator would clarify whether the strangeness enhancement observed in heavy ion collisions is indeed the result of producing a chemically equilibrated plasma. 

Other correlators are presented in \cite{Pratt:2012dz}. The $p\bar{p}$ correlator provides excellent insight into the density of up and down quarks in the QGP. In the $pK^-$ correlator the short and long-range components have opposite signs. This would provide unequivocal evidence of the two-wave nature of charge production. Charge balance functions of unidentified particles ($\alpha/\beta$ refer to all positive/negative particles), which are simply the correlator divided by the multiplicity, have been observed to narrow with centrality \cite{Aggarwal:2010ya,Adams:2003kg,Westfall:2004cq}. This narrowing is consistent with many of the up and down quarks being produced late in central collisions \cite{Bass:2000az}. To truly test this interpretation, one should observe a lack of narrowing in the both the $K^+K^-$ and $p\bar{p}$ balance functions. Although the discussion above is mainly qualitative, these correlations should provide detailed quantitative tests of the chemical evolution of the QGP. Additionally, some of the hadronic correlators are sensitive to the off-diagonal elements of $g_{ab}$, and therefore provide additional insight into whether partons move independently. Finally, through Eq. (\ref{eq:realthing}), one can test more realistic models of the chemical evolution. The advent of good time-of-flight measurements in large acceptance detectors in STAR at RHIC and in ALICE at the LHC will provide the means to apply the ideas presented here. Recent high statistics runs should provide the necessary statistics to pursue all the ideas discussed here, and finally provide strong tests of whether the quark-gluon plasma is indeed a plasma of independently moving partons with densities corresponding to equilibrium.

\acknowledgments{This research was supported by the Department of Energy's Office of Science, grant No. DE-FG02-03ER41259. The assistance of Claudia Ratti in providing the lattice data in Figure \ref{fig:claudia} is gratefully acknowledged.}

\end{document}